\documentclass[aps, pra, reprint, onecolumn, showpacs, amsmath, amssymb]{revtex4-1}

\usepackage{amsmath,amscd}
\usepackage{bm}% bold math
\usepackage[english]{babel}
\usepackage{graphicx}% Include figure files
\usepackage{soul}
\usepackage[usenames]{color}
\usepackage{subfig}
\begin{document}
\title{ Surface radiation of a charged particle bunch passing through a corrugated surface with a relatively small period }

\author{ Evgeniy S. Simakov }
\email{simakov.eugeniy@gmail.com}
\author{ Andrey V. Tyukhtin }
\affiliation{Saint Petersburg State University, 7/9 Universitetskaya nab., St. Petersburg, 199034 Russia}

\date{\today}

\begin{abstract}
	We investigate radiation of a charged particle bunch moving through a corrugated planar conductive surface. 
	It is assumed that the corrugation period and depth are much less than the wavelengths under consideration. 
	In this case, the corrugated structure can be replaced with some smooth surface on which the so-called equivalent boundary conditions (EBC) are fulfilled. 
	Using the EBC method we obtain expressions for the electromagnetic field of the bunch which are presented in form of spectral integrals. 
	It is demonstrated that the bunch generates surface waves propagating along the corrugations with the light velocity. 
	Also we present results of numerical calculations for electromagnetic field components of surface waves depending on coordinates and show that these dependences can be used for determination of the bunch size.
\end{abstract}

\maketitle

\section{Introduction\label{sec:intro}}

Methods of generation of microwave and Terahertz (THz) emission are actively developed nowadays for a number of reasons. 
These types of radiation are the subject of significant interest and find prospective applications in many fields of science and technology. 
One of the sources of microwave and THz radiation is electromagnetic interaction of charged particles with periodic structures.
Usually, researchers investigate situations where the wavelengths are comparable with the structure period.
However, there is also a number of works where relatively large wavelengths are considered.
In particular, some papers are devoted to theoretical investigations (based on surface impedance formalism) of radiation in circular metallic waveguides with fine corrugation~%
\cite{1,2,3}.
It should be noted that microwave and THz radiation were also studied in some experimental works~%
\cite{4,5}.

In this paper we consider electromagnetic radiation of a charged particle bunch passing through the corrugated surface having rectangular "cells". 
We assume that the wavelengths under consideration are much larger than the structure period and the depth of corrugation.
In this case, the solution of the problem can be obtained by using the equivalent boundary conditions (EBC) which should be fulfilled on the flat surface~%
\cite{6}. 
Thus, we analyze so-called "long-wave" radiation which differs principally from well-known Smith-Purcell radiation (SPR) where wavelengths are comparable or less than the structure period (note that SPR is well-studied physical phenomena - see, for example, the book~%
\cite{7}).
Note that the interaction of the field of the charged particle bunch  with anisotropic surface often results in generation of surface waves (see, for example,~%
\cite{8, 9}). 
In present work we will show, in particular, that this interesting phenomenon gives essential advantages for the bunch diagnostics.

The problem under consideration was not solved previously, despite numerous papers dedicated to radiation in the presence of periodic structures.
It should be mentioned that the use of the EBC method has been justified, in particular, in paper~%
\cite{3} where authors have been studied radiation from charged particle bunch moving in a circular waveguide with shallow corrugation.
The comparison between theoretical results and results of the CST Particle Studio showed good coincidence.
Paper~%
\cite{8} demonstrates analytical solution of a problem where a charged particle bunch moves along a planar surface with fine corrugation. 
Using the EBC method we obtained general solution and carried out an asymptotic analysis of the field. 
It was shown that the ultrarelativistic bunch generates the surface waves propagating along the structure, whereas the volume radiation is absent. 

The present paper is organized as follows. 
After introduction (section~%
\ref{sec:intro}), we recall the EBC method (section~%
\ref{sec:EBC}) and obtain general solution of the problem (section~%
\ref{sec:GenSol}).
Section~%
\ref{sec:SurfW} is dedicated to description of the main physical effect which consists in generation of surface waves. 
In section~%
\ref{sec:Res} we present dependences of the surface waves components on bunch parameters and discuss obtained results.

\section{\label{sec:EBC}Equivalent boundary conditions}
\begin{figure*}[htbp]
	\centering % \begin{center}/\end{center} takes some additional vertical space
	\includegraphics[width=12cm,height=4cm]{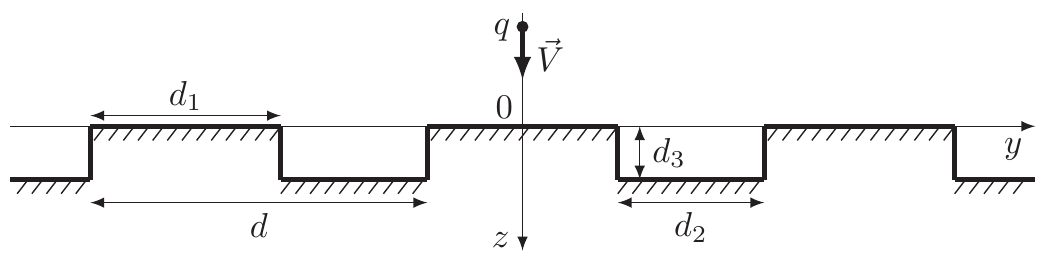}
	\caption{\label{fig:1} The corrugated surface and moving charge.}
\end{figure*}

We consider a perfectly conductive planar surface having rectangular corrugation (figure~\ref{fig:1}). 
It is assumed that the period $d$ and the depth of corrugation $d_3$ are much less than the wavelength under consideration $\lambda$: $d\ll\lambda, \; d_3\ll\lambda$. 
In this case, we can replace the corrugated surface with a flat surface on which the so-called equivalent boundary conditions (EBC) are fulfilled~%
\cite{6}.

Note that the derivation of the EBC is a complex mathematical problem involving series of cumbersome analytical calculations. 
Very briefly, the following can be said about this derivation (see the book~%
\cite{6}). 
Authors solve the problem separately for the fields of two polarizations. 
Expressions for the field components nearby the conductors are obtained from the Laplace equation with the use of conformal mapping technique. 
Lorentz lemma is used to connect the field nearby the conductors with the field in the wave zone. 
Authors come to a system of equations for the reflection and transmission coefficients. 
In fact, the EBC are determined by these coefficients. 

Note that the analogical approach was used in papers~%
\cite{1, 2} for the particular case of rectangular cells. 
Authors show that the corrugated surface can be replaced by certain thin layer with dielectric and magnetic properties.  
The surface impedance is determined by values of these constants.
However, in distinct to the EBC in~%
\cite{6}, one of the constants should be obtained numerically from the Laplace equation. 
In addition, authors consider only one polarization, and this fact does not allow us to use the conditions from~%
\cite{1, 2}.  

The EBC for Fourier-transforms of electric and magnetic fields on the plane $z=-0$ can be written in the following form~%
\cite{6}:
\begin{equation}
\label{eq:2.1}
\tag{2.1}
E_{\omega y}=\eta^m H_{\omega x}, \; E_{\omega x}=\eta^e H_{\omega y},
\end{equation}
where $\eta^m$ and $\eta^e$ are ``impedances'' which are imaginary for perfectly conductive structures. 
The boundary conditions~%
\eqref{eq:2.1} are written for the boundary problem for the area $z<0$ (for the area $z>0$ we should change the signs on right-hand sides of~%
\eqref{eq:2.1}).
Below, we will consider the area $z<0$.
In the case of the structure shown in figure~%
\ref{fig:1}, impedances $\eta^m$ and $\eta^e$ in~%
\eqref{eq:2.1} have the form~%
\cite{6}:
\begin{equation}
\tag{2.2}
\label{eq:2.2}
\eta^m=\frac{ik_0}{k_0^2-k_x^2}\left[d_0\left(k_0^2-k_x^2\right)-\delta k_y^2\right]\!,\;\;\eta^e=-i\frac{\delta}{k_0}\left(k_0^2-k_x^2\right)\!,
\end{equation}
where $d_0=d_2d_3/d$, $d_2$ is a width of groove, $k_x$ and $k_y$ are tangential components of the wave vector, $k_0=\omega/c$. 
The parameter of corrugation $\delta$ is determined by the formula~%
\cite{6}:
\begin{equation}
\label{eq:2.3}
\tag{2.3}
\delta=d_3+\frac{d}{2\pi}\ln\left(\frac{\sigma-1}{\sigma}\right)+\frac{td}{2\pi}\int_{0}^{1/\sigma}\frac{du}{\sqrt{\left(1-u\right)\left(1-\sigma u\right)}\left(\sqrt{1-tu}+1\right)},
\end{equation}
where parameters $t$ and $\sigma$ should be found from the following system of transcendent equations:
\begin{equation}
\label{eq:2.4}
\tag{2.4}
\int_{0}^{t}\frac{\sqrt{t-u}}{\sqrt{u\left(1-u\right)\left(\sigma-u\right)}}du=\pi\frac{d_1}{d},\quad\int_{t}^{1}\frac{\sqrt{u-t}}{\sqrt{u\left(1-u\right)\left(\sigma-u\right)}}du=2\pi\frac{d_3}{d}.
\end{equation}
Note that the parameter $\delta$ is of the same order as $d$ and $d_3$. 

\section{\label{sec:GenSol}General solution of the problem}

We assume that the charged particle bunch passes through the corrugated surface with velocity $\mathbf{V}=c\beta\mathbf{e}_z$ which is perpendicular to the structure. 
The charge density is written in the form $\rho=q\delta(x)\delta(y)\eta(z-Vt)$ where $\eta(z-Vt)$ is the charge distribution along the trajectory of the charge motion. 
We will use the Hertz potential $\mathbf{\Pi}$ and present it as a sum of the ``forced'' field potential $\mathbf{\Pi}^{(i)}$ and the ``free'' field potential $\mathbf{\Pi}^{(r)}$: $\mathbf{\Pi}=\mathbf{\Pi}^{(i)}+\mathbf{\Pi}^{(r)}=\int_{-\infty}^{+\infty}\left(\mathbf{\Pi}_\omega^{(i)}+\mathbf{\Pi}_\omega^{(r)}\right)e^{-i\omega t}d\omega,$
where $\mathbf{\Pi}_\omega^{(i)}$ and $\mathbf{\Pi}_\omega^{(r)}$ are corresponding Fourier-transforms. 
We mean that the ``forced'' field is the field in unbounded vacuum, and the ``free'' field is an additional field connected with the influence of the corrugated structure. 
Relations between the Fourier-transforms of electromagnetic field and Hertz potential are given by the formulas $\mathbf{E}_\omega=\mathbf{\nabla}\operatorname{div}\mathbf{\Pi}_\omega+k_0^2\mathbf{\Pi}_\omega, \; \mathbf{H}_\omega=-ik_0\operatorname{rot}\mathbf{\Pi}_\omega $. 
It is well known that from Maxwell's equations one can obtain the Helmholtz equation for the Hertz potential:
\begin{equation}
\label{eq:3.1}
\tag{3.1}
\left(\Delta+k_0^2\right)\mathbf{\Pi}_\omega=-\frac{4\pi i}{ck_0}\mathbf{j}_\omega.
\end{equation}
The solution of eq.~%
\eqref{eq:3.1} for the ``forced'' field is the well-known Coulomb field of moving charge in unbounded vacuum.
The Fourier-transforms of the Hertz vector components of this field are
\begin{equation}
\label{eq:3.2}
\tag{3.2}
\Pi_{\omega x}^{(i)}=\Pi_{\omega y}^{(i)}=0,\quad\Pi_{\omega z}^{(i)}=\frac{iq\tilde\eta}{\pi k_0c}e^{i\frac{k_0z}{\beta}}\iint_{-\infty}^{+\infty}dk_xdk_y\frac{e^{ik_xx+ik_yy}}{k_x^{2}+k_y^{2}+k_0^{2}\frac{1-\beta^2}{\beta^2}},
\end{equation}
where $\tilde\eta$ is the Fourier-transform of the charge distribution along the trajectory of the motion:
\begin{equation}
\label{eq:3.3}
\tag{3.3}
\tilde\eta=\frac{1}{2\pi}\int_{-\infty}^{+\infty}d\zeta\eta\left(\zeta\right)e^{-i\frac{k_0}{\beta}\zeta}, \;\;\zeta=z-vt.
\end{equation}

We will describe the ``free'' field with help of two-component Hertz vector $\mathbf{\Pi}_\omega^{(r)}=\Pi_{\omega x}^{(r)}\mathbf{e}_x+\Pi_{\omega z}^{(r)}\mathbf{e}_z$.
Requiring the fulfillment of the Helmholtz equation and the radiation condition (the waves must propagate from the plane $z=0$) we obtain the following expressions for the components of this Hertz vector:
\begin{align}
\label{eq:3.4}
\tag{3.4}
\left\{\begin{aligned}
\Pi_{\omega x}^{(r)} \\
\Pi_{\omega z}^{(r)}\end{aligned}\right\}
=\frac{iq\tilde\eta}{\pi k_0^{2}c}\iint_{-\infty}^{+\infty}dk_xdk_y
\left\{\begin{aligned}
R_x \\
R_z\end{aligned}\right\}
\frac{e^{ik_xx+ik_yy+ik_{z0}|z|}}{k_{z0}},
\end{align}
where $k_{z0}=\sqrt{k_0^2-k_x^2-k_y^2}$ ($k_{z0}>0$ for a positive radical expression and $\operatorname{Im}k_{z0}>0$ for a negative one). 
Here, $R_x$ and $R_z$ should be found from the boundary conditions. Substituting Fourier-transforms of total electromagnetic field components in system of the boundary conditions~%
\eqref{eq:2.1} we obtain the following expressions for them:
\begin{equation}
\label{eq:3.5}
\tag{3.5}
R_x=-\frac{\beta k_0k_xk_{z0}\left(\beta k_{z0}+k_0\right)\left(\eta^m+\eta^e\right)}{\left(k_0^2-\beta^2k_{z0}^2\right)\left[k_0k_{z0}+\left(k_x^2+k_{z0}^2\right)\eta^e-\left(k_0^2-k_x^2\right)\eta^m-k_0k_{z0}\eta^e\eta^m\right]},
\end{equation}
\begin{equation}
\label{eq:3.6}
\tag{3.6}
R_z=\frac{\beta k_0k_{z0}\left[k_0^2-\left(\beta k_x^2-k_0k_{z0}\right)\eta^e+\beta\left(k_0^2-k_x^2\right)\eta^m+\beta k_0k_{z0}\eta^e\eta^m\right]}{\left(k_0^2-\beta^2k_{z0}^2\right)\left[k_0k_{z0}+\left(k_x^2+k_{z0}^2\right)\eta^e-\left(k_0^2-k_x^2\right)\eta^m-k_0k_{z0}\eta^e\eta^m\right]}.
\end{equation}
Thus, we have found the general solution of the problem.

\section{\label{sec:SurfW}Surface waves}

The obtained results have been investigated asymptotically under the condition $k_0|x|\gg1$ using the saddle point method~%
\cite{10}. 
Note that taking into account expressions for impedances~%
\eqref{eq:2.2} one can show that $R_x$ has several peculiarities including the poles $k_x=\pm k_0$ which have the most interest for us.
These poles appear due to the impedance $\eta^m$ in the numerator of the expression for $R_x$ (see~%
\eqref{eq:2.2}).
One can show that $R_z$ has no these peculiarities.

Asymptotical investigation shows that the initial integration path can be transformed to the steepest descent path with separation of the contributions of the pole $k_x=+k_0$ for $x>0$ and $k_x=-k_0$ for $x<0$. 
These  contributions exist necessarily if the observation point is close to the corrugated structure. 
As it is shown below, these contributions are the surface waves. 

Note that the contribution of the saddle point is the volume radiation. 
We do not consider here this phenomenon because it differs from the volume radiation generated in the case of a smooth metallic surface by only the corrections of the order of small parameter $k_0d$. 
Therefore we focus further on the description of the surface waves which can not be generated for the case of a smooth (isotropic) surface. 

Contribution of the poles $k_x=\pm k_0$ can be written in the form 
\begin{align}
\label{eq:4.1}
\notag
\Pi_{z}^{(s)}\!=\!0,\quad%
\Pi_{x}^{(s)}\!=%
2\pi &i\operatorname{sgn}\left(x\right)\int_{-\infty}^{+\infty}d\omega e^{-i\omega t}\underset{k_x = \pm k_0}{\operatorname{Res}}\Pi_{\omega x}^{(r)}=2iq\beta\operatorname{sgn}\left(x\right)\delta\,\times\\
&\times\int_{-\infty}^{+\infty}dk_0\tilde\eta e^{ik_0\left(|x|-ct\right)}\int_{0}^{+\infty}\!dk_y\frac{i\beta k_y+k_0}{k_y^{2}+k_0^{2}/\beta^2}\operatorname{cos}\left(k_yy\right)e^{-k_y|z|}.
\end{align}
The corresponding field components are
\begin{align}
\label{eq:4.2}
\notag
\left\{\begin{aligned}
E_{y}^{(s)} \\
E_{z}^{(s)}\end{aligned}\right\}
=\operatorname{sgn}\left(x\right)&%
\left\{\begin{aligned}
H_{z}^{(s)} \\
-H_{y}^{(s)}\end{aligned}\right\}
=2q\beta\delta\int_{-\infty}^{+\infty}dk_0k_0\tilde\eta e^{ik_0\left(|x|-ct\right)}\times\\
\tag{4.2}
&\times\int_{0}^{+\infty}dk_y\frac{k_y\left(i\beta k_y+k_0\right)}{k_y^{2}+k_0^{2}/\beta^2}
\left\{\begin{aligned}
&\!\!\!\:\operatorname{sin}\left(k_yy\right) \\
-&\operatorname{cos}\left(k_yy\right)\end{aligned}\right\}
e^{-k_y|z|}.
\end{align}
We remind that this result is derived for the area $z<0$.
Due to the problem symmetry it is also applicable for the region $z>0$ if we take into account evenness of $E_y^{(s)}$ and oddness of $E_z^{(s)}$.

As we can see, the integrands in~%
\eqref{eq:4.2} decrease exponentially with an increase in $|z|$, and therefore describe the surface waves propagating in the plane of the corrugated structure. 
The time dependence is included only in the combination $|x|-ct$, that is, the waves propagate along the corrugation with the speed of light without attenuation due to the perfect conductivity of the structure. 
Naturally, in reality, surface waves will decay due to the finite conductivity of the metal, however, this effect is insignificant for well-conducting structures (for example, in~%
\cite{9} this is shown for a system of thin conductors). 

\section{\label{sec:Res}Numerical results}

In this section we present results of computation of surface wave component $E_z^{(s)}$ depending on values of $z$, $y$ and $|x|-ct$, according to formula~%
\eqref{eq:4.2}. 
Note that the use of the "shifted" coordinate $|x|-ct$ is connected with the fact that the surface wave propagates along $x$-axis with the light velocity $c$.
We consider Gaussian bunch with charge distribution $\eta_{gaus}\left(\zeta\right)$ and corresponding Fourier-transform $\tilde\eta_{gaus}$ determined by formulas
\begin{equation}
\label{eq:5.1}
\tag{5.1}
\eta_{gaus}\left(\zeta\right)=\frac{\exp\left(-\zeta^2/2\sigma^2\right)}{\sqrt{2\pi}\sigma}, \;\;\tilde\eta_{gaus}=\frac{\exp\left(-k_0^2\sigma^2/2\beta^2\right)}{2\pi}.
\end{equation}
Also we study "rectangular" bunch for which 
\begin{equation}
\label{eq:5.2}
\tag{5.2}
\eta_{rect}\left(\zeta\right)=\frac{\Theta\left(\sigma-|\zeta|\right)}{2\sigma}, \;\;\tilde\eta_{rect}=\frac{1}{2\pi}\frac{\sin\left(k_0\sigma/\beta\right)}{k_0\sigma/\beta}.
\end{equation}
In eqs.~%
\eqref{eq:5.1} and~%
\eqref{eq:5.2} $\sigma$ is a half of the bunch length, $\Theta\left(\sigma-|\zeta|\right)$ is the Heaviside step function.

\begin{figure}
\includegraphics[width=0.95\linewidth]{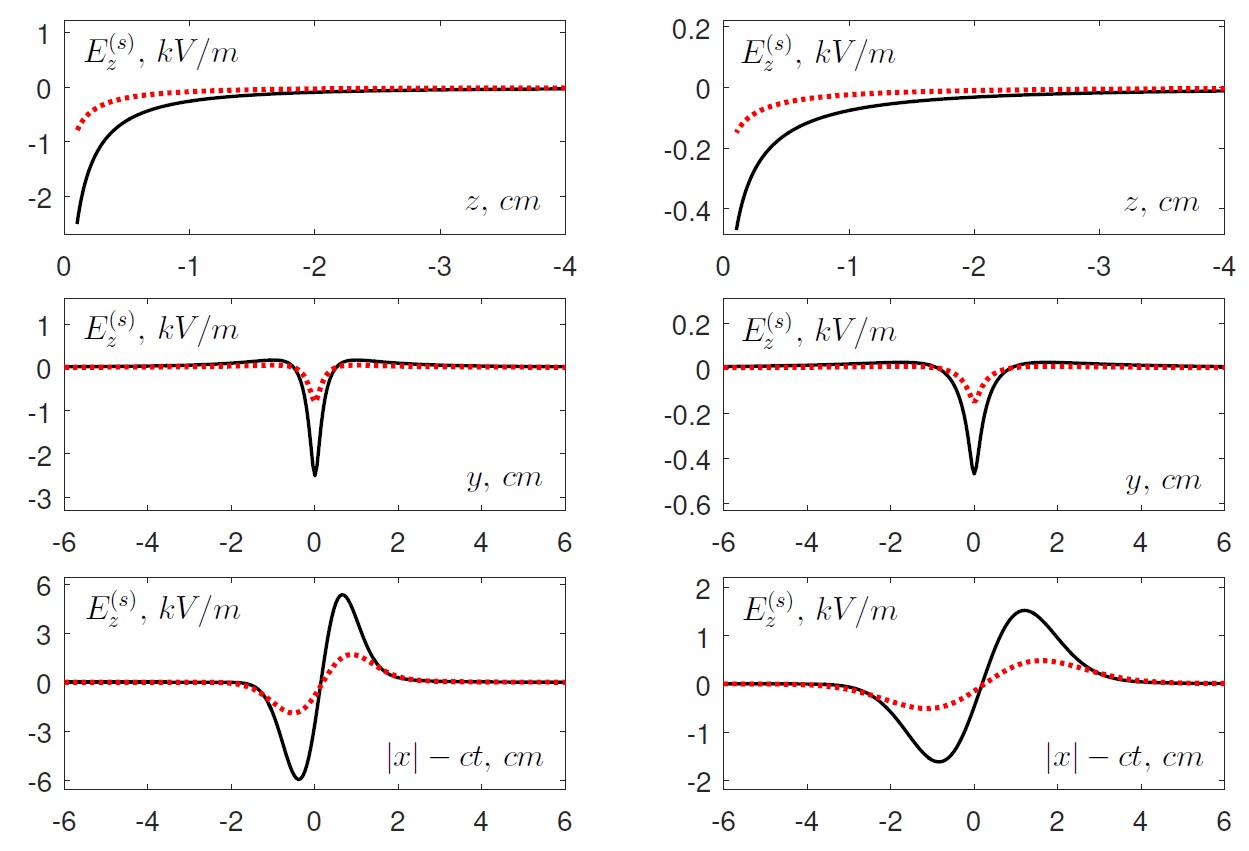}
%\begin{figure}[htb]
%	\centering
%	\subfloat{%
%		\makebox[\textwidth][c]{\includegraphics{Ez_z_gauss.eps}}} \\
%	\vspace{-11pt}
%	\subfloat{%
%		\makebox[\textwidth][c]{\includegraphics{Ez_y_gauss.eps}}} \\
%	\vspace{-8pt}
%	\subfloat{%
%		\makebox[\textwidth][c]{\includegraphics{Ez_x_gauss.eps}}} \\
\caption{\label{fig:2} The component of the surface wave $E_z$ depending on the coordinates $z$ (top row), $y$ (middle row) and $|x|-ct$ (bottom row) for the Gaussian bunch with $q=+1$ nC. 
	The relative velocity of the bunch is $\beta=1$ (solid black curves) and $\beta=0.75$ (dotted red curves). 
	The bunch length is $2\sigma=1$ cm (left coloumn) and $2\sigma=2$ cm (right coloumn). 
	The period and the depth of the structure are $d=d_3=0.1$ cm, the width of "hills" is $d_1=0.05$ cm. The values of coordinates are $y=0$ cm, $|x|=ct$ (top row); $z=-0.1$ cm, $|x|=ct$ (middle row); $z=-0.1$ cm, $y=0$ cm (bottom row).}
\end{figure}

\begin{figure}[htb]
\includegraphics[width=0.95\linewidth]{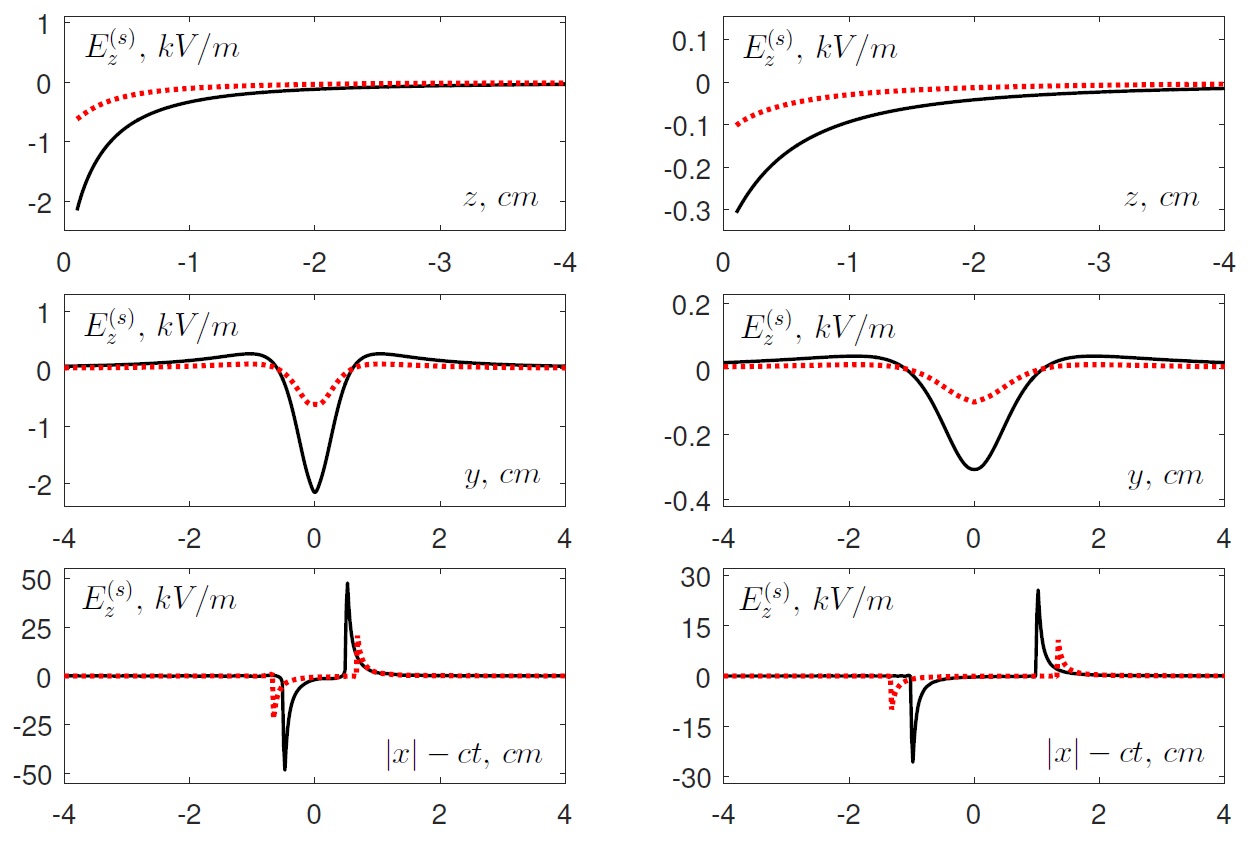}
%	\centering
%	\subfloat{%
%		\makebox[\textwidth][c]{\includegraphics{Ez_z_rect.eps}}} \\
%	\vspace{-11pt}
%	\subfloat{%
%		\makebox[\textwidth][c]{\includegraphics{Ez_y_rect.eps}}} \\
%	\vspace{-6pt}
%	\subfloat{%
%		\makebox[\textwidth][c]{\includegraphics{Ez_x_rect.eps}}} \\
\caption{\label{fig:3} The same as in figure~\ref{fig:2} for the "rectangular" bunch.}
\end{figure}

In figures~%
\ref{fig:2} and~%
\ref{fig:3} one can see that the component $E_z^{(s)}$ decreases rapidly with increasing in $|z|$ (plots in top rows) and have the sharp extremum at $y=0$ for any time (plots in middle rows). 
Dependences of the surface wave component $E_z^{(s)}$ on "shifted" coordinate $|x|-ct$ (plots in bottom rows) show that they allow determining the length of the bunch, especially in the case of the "rectangular" bunch (figure~%
\ref{fig:3}):
the distance between maximums is equal to the bunch length divided by $\beta$.
In addition, figures~%
\ref{fig:2} and~%
\ref{fig:3} show that the field value significantly decreases with increasing in bunch length $2\sigma$.
It is also important to note that the field magnitude decreases with a decrease in velocity $\beta$, but there is no the velocity threshold below which the surface waves are not generated (in contrast to the problem solved in paper~%
\cite{8} where the surface radiation is excited by ultrarelativistic bunch only).

In conclusion, we briefly discuss the energy of the surface waves. 
The estimation shows that the total spectral density of the surface wave energy is much less than the one of the volume radiation (approximately in $k_0^2 d^2$ times). 
However, in contrast to the volume radiation, the surface waves are concentrated in certain small area nearby the corrugated surface.
For instance, in situations shown in figures~%
\ref{fig:2} 
and~%
\ref{fig:3} 
the majority of the surface wave energy passes through the square of the order of $1\,\text{cm}^2$ in the plane $(y,z)$. 
This fact allows to hope for relatively simple detection of surface waves in experiment. 

\section{Acknowledgements}

This research was supported by the Russian Science Foundation, Grant No. 18-72-10137.

\end{document}